# Electromagnetic, stress and thermal analysis of the Superconducting Magnet

Yong Ren[1]*, Xiang Gao[1]*, [2]

[1] Institute of Plasma Physics, Hefei Institutes of Physical Science, Chinese Academy of Sciences, PO Box 1126, Hefei, Anhui, 230031, People's Republic of China

[2] University of Science and Technology of China, Hefei, Anhui 230026, People's Republic of China

*E-mail: renyong@mail.ustc.edu.cn and xgao@ipp.ac.cn

*Abstract*—**Within the framework of the National Special Project for Magnetic Confined Nuclear Fusion Energy of China, the design of a superconducting magnet project as a test facility of the Nb3Sn coil or NbTi coil for the Chinese Fusion Engineering Test Reactor (CFETR) has been carried out not only to estimate the relevant conductor performance but also to implement a background magnetic field for CFETR CS insert and toroidal field (TF) insert coils. The superconducting magnet is composed of two parts: the inner part with Nb3Sn cable-in-conduit conductor (CICC) and the outer part with NbTi CICC. Both parts are connected in series and powered by a single DC power supply. The superconducting magnet can be cooled with supercritical helium at inlet temperature of 4.5 K. The total inductance and stored energy of the superconducting magnet are about 0.278 H and 436.6 MJ at an operating current of 56 kA respectively. An active quench protection circuit was adopted to transfer the stored magnetic energy of the superconducting magnet during a dump operation.**

**In this paper, the design of the superconducting magnet and the main analysis results of the electromagnetic, structural and thermal-hydraulic performance are described.**

## I. INTRODUCTION

CHINESE Fusion Engineering Test Reactor (CFETR) is being designed to bridge the gaps between the ITER and Demo in China [1-6]. The superconducting magnet for the CFETR reactor consists of a CS coil with 6 modules, 8 PF coils, 16 TF coils and a set of correction coils [6]. The CFETR CS coil with Nb3Sn conductor will operate in pulsed mode. The electromagnetic and thermal cyclic operation often results in Nb3Sn CICC conductor performance degradation [7-19]. Generally, the qualification tests of the Nb3Sn CICC conductor in electromagnetic and thermal cyclic operation were performed to evaluate the relevant Nb3Sn CICC coil performance [14-19]. Occasionally, the qualification tests of the Nb3Sn CICC conductors with a long length in relevant conditions of magnetic field, current density and mechanical strain are of great importance for the fabrication of the large scale superconducting magnet [20-33]. Therefore, a design activity has been started to design a superconducting magnet project not only to estimate the CFETR CS coil performance but also to implement a background field superconducting magnet for testing the CFETR CS insert and TF insert coils. During the first stage, the main goals of the project are composed of: 1) to obtain the maximum magnetic field of above 12.5 T; 2) to simulate the relevant thermal-hydraulic characteristics of the CFETR CS coil; 3) to test the sensitivity of the current sharing temperature to electromagnetic and thermal cyclic operation. During the second stage, the superconducting magnet was used to test the CFETR CS insert and TF insert coils as a background magnetic field superconducting magnet during the manufacturing stage of the CFETR magnets.

In this paper, the magnetic field, the strain of the Nb3Sn cable and the stress of the jacket for the superconducting magnet with a detailed 2-D finite element method were analyzed. The temperature margin behavior of the superconducting magnet was analyzed with the 1-D GANDALF code [34]. The quench behavior of the superconducting magnet was given with the adiabatic hot spot temperature criterion and the 1-D GANDALF code.

## II. DESIGN OF THE SUPERCONDUCTING MAGNET

The superconducting magnet is designed to provide a maximum magnetic field of above 12.5 T. The superconducting magnet should have enough inner space to provide a background magnetic field for testing the long superconducting samples such as CFETR CS insert and TF insert coils. The Nb3Sn cable-in-conduit conductor (CICC) and NbTi CICC are graded for reducing the cost of the superconducting strands for the superconducting magnet. The superconducting magnet, which has a cold bore of 1400 mm in diameter, will produce about 12.6 T maximum magnetic field at 56 kA. The total self-inductance and stored energy of the superconducting magnet are about 0.278 H and 436.6 MJ respectively. Table 1 lists the main parameters of the superconducting magnet. Figure 1 shows the cross section of a winding pack of the superconducting magnet. Figure 2 shows the magnetic field distribution of the superconducting magnet.

The superconducting magnet consists of two modules. The inner module is layer-wound winding; the outer module is pancake-winding. The inner module with Nb3Sn CICC has eight layers; each layer has one cooling channel. The outer module with NbTi CICC has 20 pancakes with 200 turns.

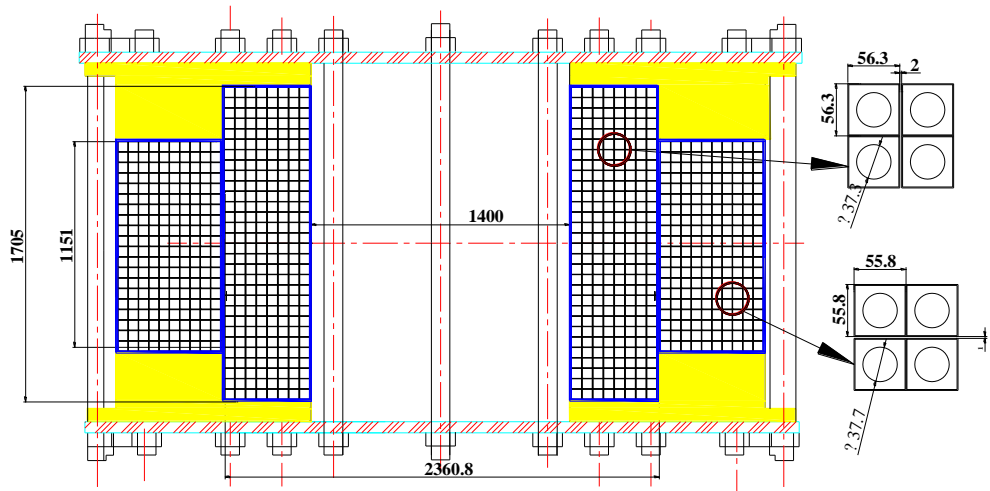

**Figure 1.** Cross section of the winding pack of the superconducting magnet.

There are 10 cooling channels for the outer module in total; each cooling channel consists of 2 pancakes. The CFETR CS type conductor was used as the inner module for the superconducting magnet, while the ITER PF6 type conductor with NbTi strand was used as the outer module [6, 35].

The round-in-square type jackets are used for both conductors. The 316LN stainless steel and 316L stainless steel jackets are adopted for the Nb₃Sn and NbTi CICC conductors. The jacket for the Nb₃Sn CICC conductor is 56.3 mm wide and has a circular hole of 37.3 mm. The jacket for the NbTi CICC conductor is 55.8 mm wide and has a circular hole of 37.7 mm [35]. Both conductors are wrapped with 1 mm thick insulation. The thicknesses of the ineterlayer insulation for the inner module and of the interturn insulation of the outer module are 1.5 mm. The ground insulation is 8 mm. Table 2 lists the specification of the Nb₃Sn and NbTi conductors [3, 4].

Unlike the distributed barrier Nb₃Sn internal tin (IT) superconducting strand used for the hybrid magnet superconducting outsert magnet, the single barrier Nb₃Sn superconducting strand, which has the highly critical current density and low AC losses during ramping, was adopted for the Nb₃Sn coil of the superconducting magnet [36-38]. The Nb₃Sn strand, with a diameter of 0.81 mm, is Cr plated to reduce the coupling loss of the superconducting cable and to prevent the diffusion reaction bonding of the strands during the reaction heat treatment process [39-48]. The NbTi strand is Ni plated to prevent gradual oxidation [49]. Table 3 lists the specification of the Nb₃Sn and NbTi strands.

**Table 1.** Design parameters of the superconducting magnet.

| Superconductor | | Nb₃Sn | NbTi |
|---|---|---|---|
| Jacket | | 316LN | 316L |
| Inner diameter | mm | 1.4000 | 2.3808 |
| Outer diameter | mm | 2.3608 | 3.5288 |
| Height | mm | 1.7050 | 1.1510 |
| Turn insulation | mm | 1.0 | 1.0 |
| Layer/pancake insulation | mm | 2.0 | 1.0 |
| Layer/Pancake | | 8 | 20 |
| Turns per layer or pancake | | 30 | 10 |
| Current | kA | | 56 |
| Inductance | H | | 0.278 |
| Stored energy | MJ | | 436.6 |
| Maximum field | T | 12.59 | 5.455 |

**Table 2.** Main parameters of the Nb₃Sn and NbTi conductors [3, 4].

| | | HF | LF |
|---|---|---|---|
| Superconductor | | Nb₃Sn | NbTi |
| Cable pattern | | (2sc + 1Cu) x 3 x 4 x6 x 6 | 3sc x 4 x 4 x 5 x 6 |
| Central spiral | mm | ID 8, OD 10 | 10 x 12 |
| Petal wrap | mm | 0.05mm thick, 70% cover | 0.05 mm thick, 50% cover |
| Cable wrap | | 0.08 mm thick, 40% overlap | 0.10 mm thick, 40% overlap |
| Cr coated strand diameter | mm | 0.81 | 0.73 |
| Cu/Non-Cu for SC | | 1.0 | 1.6 |
| Void fraction | | 0.30 | 0.343 |
| Cable diameter | mm | 37.3 | 37.7 |
| Conductor dimension | mm | 56.3 ×56.3 | 53.8 ×53.8 |
| Jacket | | Circle in square 316LN | Circle in square 316L |

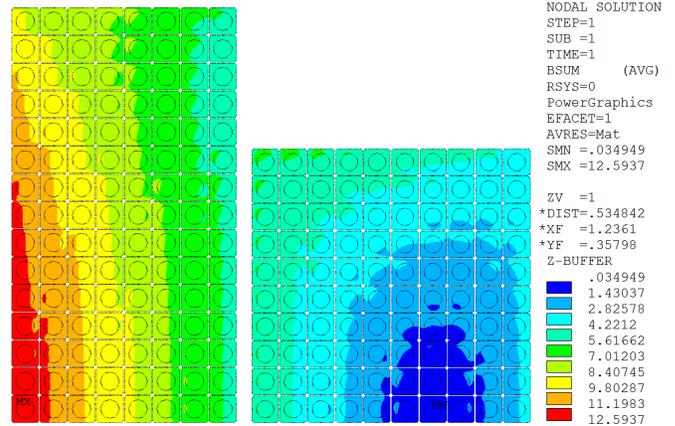

**Figure 2.** Magnetic field distribution of the superconducting magnet at 56 kA (Unit: T).

**Table 3.** Specification of the Nb₃Sn and NbTi strands used for the superconducting magnet.

| Superconductor | | Nb₃Sn | NbTi |
|---|---|---|---|
| Minimum piece length | m | 1000 | 1000 |
| Strand diameter | mm | 0.81 | 0.73 |
| Twist pitch | mm | 15 ± 2 | 15 ± 2 |
| Strand coating | μm | 2 (Cr) | 2 (Ni) |
| Cu/Non-Cu | | 1 | 1.6 |
| Critical current at 4.2 K | A | >250@12 T | >300@6.6 T |
| n value | | >20 | >20 |
| Residual resistivity | | >100 | >100 |
| ratio | | | |
| Hysteresis loss per strand unit volume | mJ/cm$^3$ | <450@4.2 K over a ± 3 T cycle | <45@4.2 K over a ± 1.5 T cycle |
| Effective filament diameter | µm | <30 | <8 |

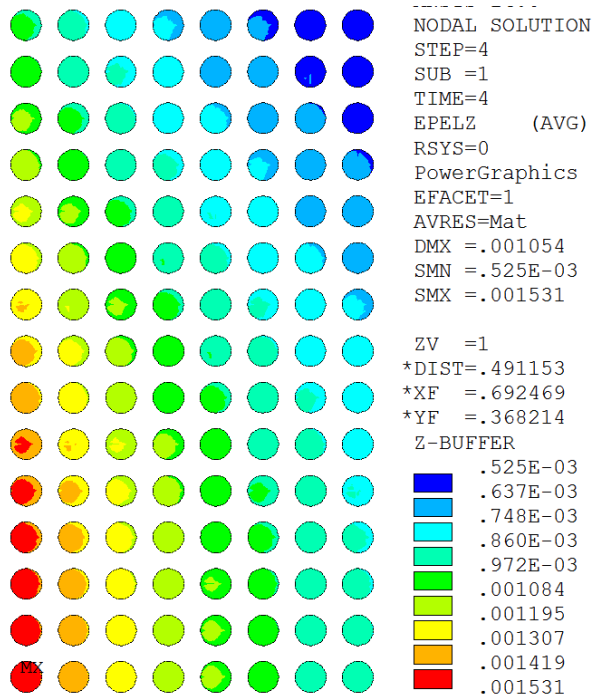

**Figure 3**. Hoop strain distribution of the superconducting cables at 56 kA.

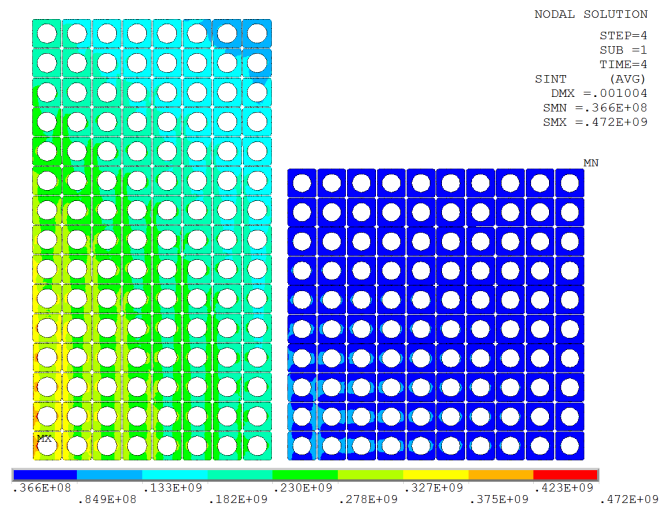

**Figure 4**. Tresca stress distribution in the jacket of the superconducting magnet (Unit: Pa).

## III. MECHANICAL BEHAVIOR ANALYSIS OF THE SUPERCONDUCTING MAGNET

The superconducting magnet will carry a large current and produce a large electromagnetic force during operation, leading to the conductor experience a large stress. The mechanical behavior needs to be analyzed to evaluate the structural integrity. The ITER structural design criteria of the superconducting magnet have been presented in [35, 49, 50]. An axis-symmetric finite element model including all components such as the jacket, cable and insulation was used to analyze the stress/strain distribution of the superconducting magnet under static operation mode with full design current applied. The cables of the superconducting magnet are assumed to be fully bonded to the inner surface of the jacket [36, 51]. The calculation are based on the combinations of the following loading conditions: the thermal contraction during cool-down, the helium pressure and the Lorentz force [36, 51].

Figure 3 shows the operating strain of the Nb$_3$Sn cable for the superconducting magnet caused by the magnetic loading. It is shown that the maximum operating strain of the Nb$_3$Sn cable from the magnetic loading is about 0.1531%.

The stress of the jacket was calculated based on the three loading conditions mentioned above. The helium pressure inside the 316LN jacket and 316L jacket during normal operations was assumed to be 0.55 MPa. The highest loaded turns of the superconducting magnet are located in the innermost turns of the Nb$_3$Sn coils. Figure 4 shows the stress distribution of the 316LN and 316L jackets of the superconducting magnet. The maximum Tresca stress value of the 316LN stainless steel is about 472 MPa. The shear stress in the insulation layer of the conductor is 12.8 MPa, which is well below the limit of 50 MPa. Figure 5 shows the shear stress distribution in the insulation layer of the superconducting magnet. The analysis results of the mechanical behavior for the superconducting magnet satisfy the ITER design criteria.

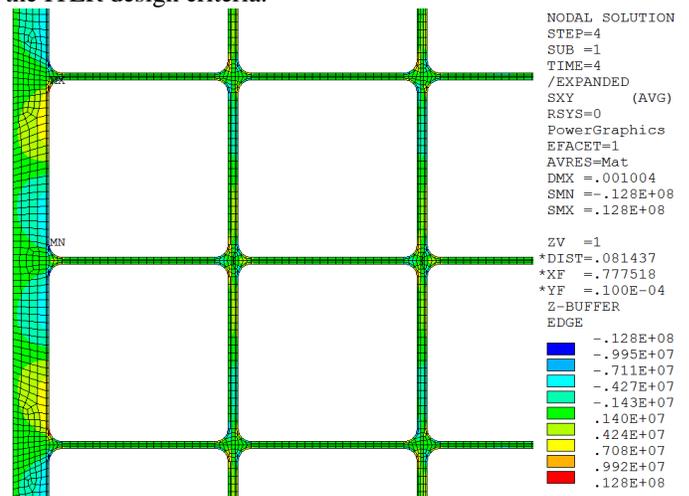

**Figure 5**. Shear stress distribution in the insulation layer of the superconducting magnet (Unit: Pa).

**Table 4.** Torque between different coils with parallel axes (Unit: N*m)

|  | Δc(mm) | 0.5 | 1.0 | 1.5 | 2.0 | 2.5 | 3.0 |
|---|---|---|---|---|---|---|---|
| Δd(mm) | 0.5 | -48.2 | -96.3 | -144.5 | -192.7 | -240.8 | -289.0 |
|  | 1.0 | -96.3 | -192.7 | -289.0 | -385.4 | -481.7 | -578.0 |
|  | 1.5 | -144.5 | -289.0 | -433.5 | -578.0 | -722.5 | -867.1 |
|  | 2.0 | -192.7 | -385.4 | -578.0 | -770.7 | -963.4 | -1156.1 |
|  | 2.5 | -240.8 | -481.7 | -722.5 | -963.4 | -1204.2 | -1445.1 |
|  | 3.0 | -289.0 | -578.0 | -867.1 | -1156.1 | -1445.1 | -1734.1 |

## IV. Electromagnetic Force And Torque due to Assembly Tolerance

Positioning deviations may occur during the assembly process. The assembly tolerance can be generally classified into the following patterns: (1) axial offset; (2) radial offset; (3) angular misalignment; and (4) the random combinations of the three patterns mentioned above [6, 52-54]. These deviations may produce a large electromagnetic force and torque. There exist a large magnetic force in axial direction for axial offset or radial force for radial offset. The axial magnetic force stiffness between the Nb₃Sn coil and the NbTi coil is about 1.4561e8 N/m where the distance is the axial displacement between coil mid-planes. The radial magnetic force between the Nb₃Sn coil and the NbTi coil is about 7.2804e4 N/m where the distance is the radial displacement between coil axes. In case of the angular offset between two inclined coils or the combinations of the axial offset and radial offset, the torque will be produced. For the inclined coils, the torque stiffness is about 5.8861e4 Nm/rad between two coils with inclined axes where rad is the angular difference between coil axes. In fact, the torque can be generated for the combination of the radial and axial offsets. Table 4 lists the torque between the Nb₃Sn coil and the NbTi coil for different radial and axial offsets. The terms Δc and Δd are denoted as axial distance between coil mid-plane and radial distance between coil axes.

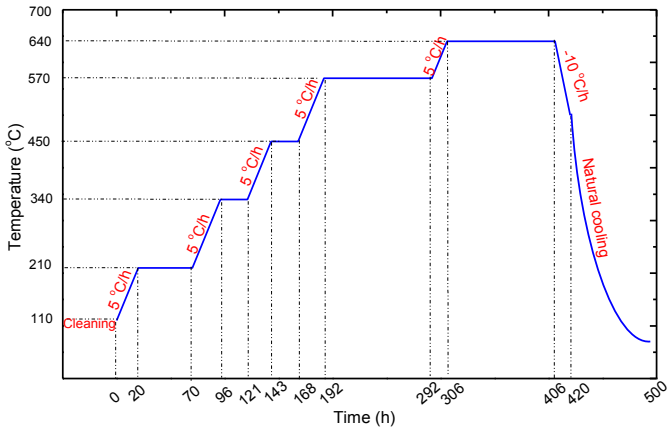

**Figure 6.** Typical reaction heat treatment process of the Nb₃Sn coil.

## V. Reaction Heat Treatment System of the Superconducting Magnet

The Nb₃Sn superconducting strands with high field performance combined with low AC losses will be adopted. The Nb₃Sn superconducting strands and copper strands are cabled together and inserted inside a 316LN jacket to form the Nb₃Sn CICC [6]. Due to the brittleness of the Nb₃Sn material and its inherent sensitivity to the mechanical forces, the wind and react technique are usually adopted to form the Nb₃Sn superconducting A15 phase [55-57]. Once the Nb₃Sn coil winding is completed, the reaction heat treatment of the Nb₃Sn coil as one of the important steps should be taken.

The microstructure and superconducting performance of the Nb₃Sn superconductors are closely related to the heat treatment temperature, ramp rate and the duration of the high temperature plateau [55-61]. The heat treatment temperature and the duration determine the grain size, the thickness of the superconducting layer and the Sn content [55, 58-60]. On one hand, the structural material and insulation material limit the maximum heat treatment temperature [61]. On the other hand, the higher heat treatment temperature until the Nb filaments are fully reacted with the Sn content, which increased the critical current density because of producing a more stoichiometric A15 layer, often decreases residual resistivity ratio (RRR) because the plated chromium on the strand diffuses to copper [58, 60]. In addition, it is widely accepted that RRR values lower with increased heat treatment time at final heat treatment plateau. Therefore, there is a fundamental compromise to be made between maximizing the grain boundary density to generate a high critical current density at high magnetic field, minimizing the AC losses and maintaining the RRR of the Cu stabilizer above 100 after the reaction heat treatment [59]. The reaction heat treatment for the Nb₃Sn coil is more complicated than that of the Nb₃Sn strands because of the complex geometry of the coil. To obtain these goals mentioned above, the maximum heat treatment temperature does not exceed 650 ℃ for the present design. The heat treatment time for the final heat treatment plateau should not exceed 150 hours. The operating temperature of the furnace can be continuously monitored with thermocouples during the reaction heat treatment period.

During the reaction heat treatment, some impurities, such as H₂O, CH and O₂ can be generated. However, once these impurities exceed a certain threshold, the jacket surface oxidation may mitigate the vacuum pressure impregnation quality and the bonding force between the epoxy-stainless steel, and then influence the mechanical performance [60]. Therefore, these impurities should be controlled in a limited range by using the purifying system. The maximum allowable value can be limited as: H₂O<10 ppm, CH<2 ppm and O₂<10 ppm. The high purity argon with 99.9995% will be supplied with two lines, one for cleaning the furnace volume, and the other one for cleaning the CICC channel to reduce the risk of contamination between the superconducting cables. The heat treatment of the Nb₃Sn coil can be implemented in vacuum or argon atmosphere. For both cases, the Nb₃Sn CICC channel should be entered the high purity argon with 99.9995% during the coil reaction heat treatment. The high purity argon gas must be maintained a positive pressure in the CICC channel. To prevent the oil enters the vacuum of the furnace cavity from the diffusion pump, the Freon diffusion pump cold traps will be installed in the exhaust port.

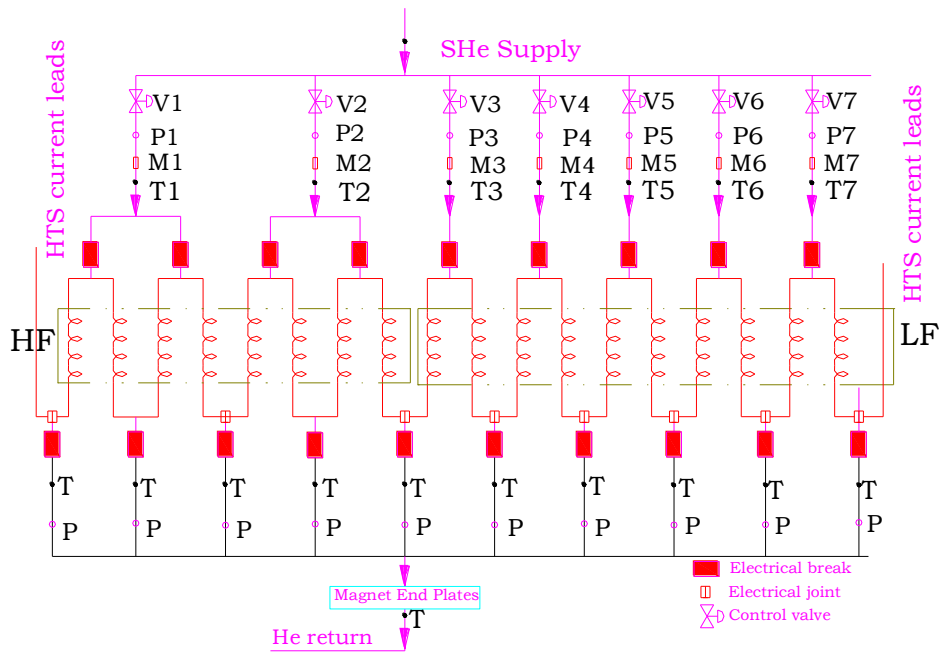

**Figure 7**. Cooling circuit of the superconducting magnet.

### A. Heat Treatment Procedure of the Nb₃Sn Coil

The nominal heat treatment schedule designed of the Nb$_3$Sn coil is shown in Figure 6. The reaction heat treatment will be carried out under vacuum as follows: 210 ℃/50 h+340℃/25 h +450℃/25 h+570 ℃/100 h+640 ℃/100 h. The maximum heating rate of the ramp up is limited to 5~10 ℃/hours to reduce the temperature difference among the whole windings, whereas the rate of the ramp down to the room temperature could be faster. It will take about 500 hours to heat the Nb$_3$Sn coil to form the Nb$_3$Sn superconductors. There are several heating elements inside the furnace region to obtain the requested uniform temperature. The allowable maximum temperature difference among the whole windings should not exceed ±5 ℃ during the dwelling period [62]. The Cu-Sn intermetallic is first formed at the low heat treatment temperature [56]. It is noted that the winding temperature should not exceed 213 ℃ during the first temperature plateau to avoid formulation of liquid tin and thus loss of tin.

The heat treatment scheme under vacuum offers two advantages: 1) effectively reduce the argon consumption; 2) ease of regulating the temperature uniformity. However, it will also present two disadvantages: 1) increases the possibility of oil contamination from the diffusion pump; 2) increases the malfunction possibility of the vacuum pump.

The heat treatment furnace can be cleaned by using argon gas for several times. The high purity argon with 99.9995% will be entered into the CICC channel with a positive pressure. At first, a large flow rate of the high purity argon was entered into the CICC channel due to the large impurities. The mass flow rate of the high purity argon can be adjusted according to the impurity concentration.

As a matter of fact, the deviation on the Nb$_3$Sn coil reaction heat treatment from the designed schedule may occur. Due to the relatively large size of the Nb$_3$Sn coil and its small thermal diffusivity, the thermal uniformity throughout the whole windings may exceed the design criteria. If the temperature difference exceed the prescribed value, the dwelling time needs to be prolonged. In addition, if the de-ionized water cooling system malfunctions, the tap water could be used as a substitute for cooling the heat furnace system. Finally, if the malfunction of the vacuum pump occurs, the heat treatment under argon atmosphere should be adopted instead of under vacuum.

### B. Emergency Actions and Remedial Actions

During the reaction heat treatment, the severe malfunction may be taken place. The malfunction consists of: 1) the malfunction of the air tightness; 2) power failure. Once the severe malfunction was taken place and cannot be repaired in a short time, the reaction heat treatment should be shut down as soon as possible and the current operation status should be documented. In the event of the malfunction of the air tightness or the power failure, the vacuum and heat will be lost. The dry purity argon should be entered into the furnace cavity to protect the Nb$_3$Sn coil from oxidation. Once the malfunction was repaired, it is necessary to heat the coil as soon as possible to the documented state and to continue the next heat treatment steps.

**Table 5.** Hydraulic parameters of the CICC conductor for the superconducting magnet.

|  |  |  | Nb₃Sn | NbTi |
|---|---|---|---|---|
| Superconductor cross section | $A_{SC}$ | mm² | 222.61 | 229.27 |
| Copper cross section | $A_{CU}$ | mm² | 445.22 | 366.80 |
| Insulation cross section | $A_{IN}$ | mm² | 450.40 | 446.40 |
| Helium cross section in bundle region | $A_{HeB}$ | mm² | 286.21 | 344.43 |
| Helium cross section in central hole | $A_{HeH}$ | mm² | 78.54 | 113.10 |
| SS jacket cross section | $A_{SS}$ | mm² | 1855.77 | 1778.16 |
| Void fraction | $V_f$ | - | 0.30 | 0.343 |
| n-index | | - | 10 | 10 |
| Residual resistivity ratio | RRR | - | 100 | 100 |
| Surface perforation from central channel to bundle | PERF0R | -- | 0.1 | 0.15 |
| Conductor Helium wetted perimeter | PHTC | mm | 2748.26 | 2752.03 |
| Jacket Helium wetted perimeter | PHTJ | mm | 58.59 | 59.22 |
| Jacket-Conductor wetted perimeter | PHTCJ | mm | 58.59 | 59.22 |
| Hole and bundle wetted perimeter | PHTHB | mm | 31.42 | 37.70 |
| Hydraulic diameter in bundle region | $D_{HB}$ | mm | 0.3995 | 0.480 |
| Hydraulic diameter in hole region | $D_{HH}$ | mm | 10.0 | 12.0 |

**Table 6.** Strand characterization parameters of the Nb₃Sn superconductors [6, 35, 63].

| Parameter | | |
|---|---|---|
| Ca1 | 44.16 | |
| Ca2 | 6.742 | |
| Eps_0a | 0.2804% | |
| Bc2m(0) | 31.75 | T |
| Tcm | 16.23 | K |
| C | 46636 | A*T |
| P | 0.9419 | |
| Q | 2.539 | |

**Table 7.** Strand characterization parameters of the NbTi superconductors [64, 65].

| Parameter | | |
|---|---|---|
| $C_0$ | 1.68512e11 | A*T |
| $Bc20$ | 14.61 | T |
| $Tc0$ | 9.03 | K |
| $\alpha$ | 1.00 | |
| $\beta$ | 1.54 | |
| $\gamma$ | 2.10 | |

## VI. Thermal-hydraulic Analysis of the Superconducting Magnet

The Nb₃Sn coils are layer-wound windings with eight layers; each layer has one cooling channel. The NbTi coils are pancake-winding with 20 pancakes. There are 10 cooling channels for NbTi coils; each cooling channel consists of two pancakes. Figure 7 shows the cooling circuit of the superconducting magnet. The Nb₃Sn CICC conductors are cooled with the forced flow supercritical helium at 0.55 MPa pressure, 12 g/s mass flow rate and 4.5 K temperature at the coil inlet. The NbTi coils are cooled with supercritical helium at inlet pressure of 0.55 MPa, mass flow rate of 8 g/s and 4.5 K temperature at the inlet. The hydraulic parameters of the Nb₃Sn and NbTi CICCs for the superconducting magnet are shown in table 5 [6, 47].

### A. Critical Current Scaling Law for the Nb₃Sn and NbTi Strands

The scaling law for the strain dependence of the critical current density in Nb₃Sn superconductor can be obtained from the relevant expressions [63]. The scaling parameters of the Nb₃Sn superconductor is shown in table 6 [6, 47, 63]. The effective filament diameter of 30 μm for the Nb₃Sn strand [6, 47]. The thermal strain of Nb₃Sn strand inside the 316LN stainless steel was assumed as -0.664 % [47]. The strain generated by the electromagnetic force was shown in figure 3. The critical current density of the NbTi superconductor can be obtained by using the single pinning model [64]. The critical current density of the NbTi superconductor can be expressed as [64, 65],

$$J_c(B,T) = \frac{C_0}{B} \left(\frac{B}{B_{C2}(T)}\right)^{\alpha} \left(1 - \frac{B}{B_{C2}(T)}\right)^{\beta} \left(1 - \left(\frac{T}{T_{C0}}\right)^{1.7}\right)^{\gamma} \quad (1)$$

$$B_{C2}(T) = B_{C20} \left(1 - \left(\frac{T}{T_{C0}}\right)^{1.7}\right)^{\gamma} \quad (2)$$

The relevant scaling law parameters of the NbTi strands can be shown in table 7. The effective filament diameter is about 8 μm for the NbTi strand. An accurate evaluation of the coupling time constant of the CICC conductors is a hard work. The coupling time constants of the CICC conductors depend on the local magnetic forces, the load cycle history, void fraction, cable pattern, aspect ratio, coating material of the cable, and the magnet ramp rate, etc. [66-71]. For simplicity, the coupling time constants of the Nb₃Sn CICC conductor and the NbTi CICC conductor were selected as 0.075 s and 0.15 s for evaluating AC losses respectively.

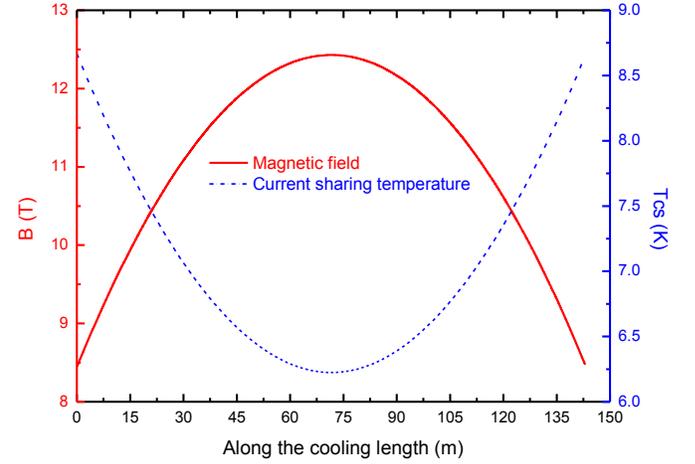

**Figure 8**. Magnetic field and current sharing temperature at 56 kA along the cooling length of the A1 channel.

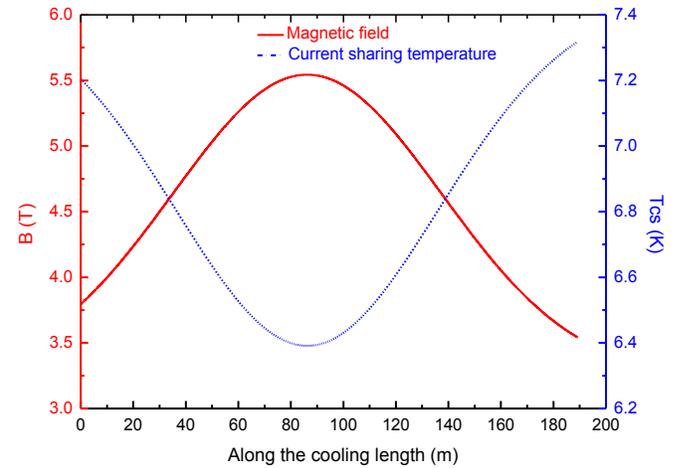

**Figure 9**. Magnetic field and current sharing temperature at 56 kA along the cooling length of the B1 channel.

### B. Friction Factor of the CICC Conductors for the Superconducting Magnet

The Nb₃Sn and NbTi coils are forced-flow cooled with supercritical helium. The pressure drop in the CICC for the central channel and the bundle of the superconducting magnet can be expressed as (3)-(5) [65, 72, 75],

$$\frac{dp}{dx} = -2\rho \frac{f}{D_h} v|v| \tag{3}$$

$$f_{bundle} = \frac{1}{4v_f^{0.742}} \left( \frac{19.5}{Re^{0.7953}} + 0.0231 \right) \tag{4}$$

$$f_{central} = 0.36 \times \frac{1}{Re^{0.04}} \times 0.25 \tag{5}$$

where $dp/dx$ is the pressure gradient, $f$ is the friction factor, $f_{bundle}$ and $f_{central}$ are friction factors of the bundle and the central channel, $\rho$ is the density of the helium, $v$ is the flow speed, $v_f$ is the void fraction, Re is the Reynolds number, $D_h$ is the hydraulic diameter.

## C. Current sharing temperature of the superconducting magnet

The thermal-hydraulic behavior of the innermost layer of the Nb$_3$Sn coil was analyzed due to its lower minimum temperature margin compared with that of the other layers for the Nb$_3$Sn coils. For the NbTi coils, the lowest value of the minimum temperature margin is located at the top and bottom channels. So, we only analyze the thermal-hydraulic behavior of the top channel of the NbTi coil. The innermost layer of the Nb$_3$Sn coil and the top channel of the NbTi coil can be referred as A1 and B1 channels. Figures 8 and 9 show the current sharing temperature of the A1 channel and the B1 channel when the superconducting magnet is ramped up to the full field. It is shown that the minimum current sharing temperature is about 6.3 K for the A1 channel and 6.4 K for the B1 channel.

One of the most important operational scenarios for the superconducting magnet is a cyclic operation when the superconducting magnet is linearly ramped up to the full field and then ramped down to zero field, cycle after cycle. The ramp rate of 280 A/s was firstly adopted in this case. Figures 10 and 11 show the maximum cable temperature and minimum temperature margin evolution as functions of time for the A1 and B1 channels. The analysis results are shown that the lowest values of the minimum temperature margin for the A1 and B1 channels are 1.50 and 1.70 respectively. Figures 12 and 13 show the maximum cable temperature and outlet temperature evolution as functions of time for the A1 and B1 channels. The results show the evolution of the outlet temperature as a function of time are nearly consistent with that of the maximum temperature for the A1 and B1 channels.

The parametric analysis is performed to evaluate the sensitivity of the minimum temperature margin for the A1 and B1 channels to the current ramp rate. Figure 14 shows the minimum temperature margin as a function of the current ramp rate for cyclic operation. It is shown that the continuous cyclic operations can be allowed for the current ramp rate below 280 A/s. With the increasing current ramp rate, the minimum temperature will drop below 1.0 K for the current ramp rate of 500 A/s.

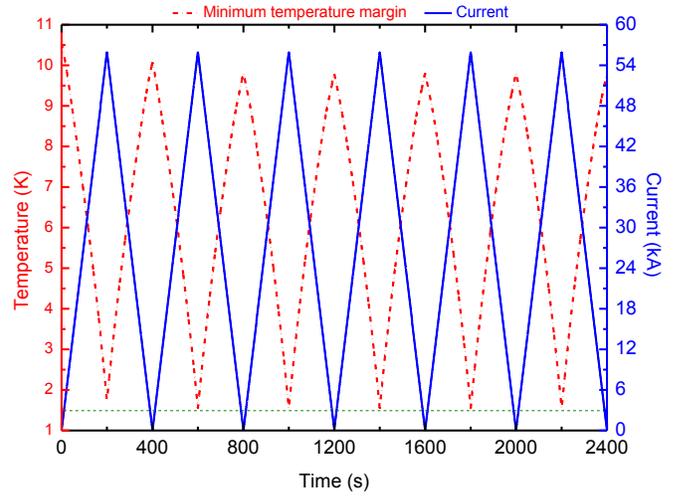

**Figure 10.** Minimum temperature margin and current evolution as functions of time for the A1 channel.

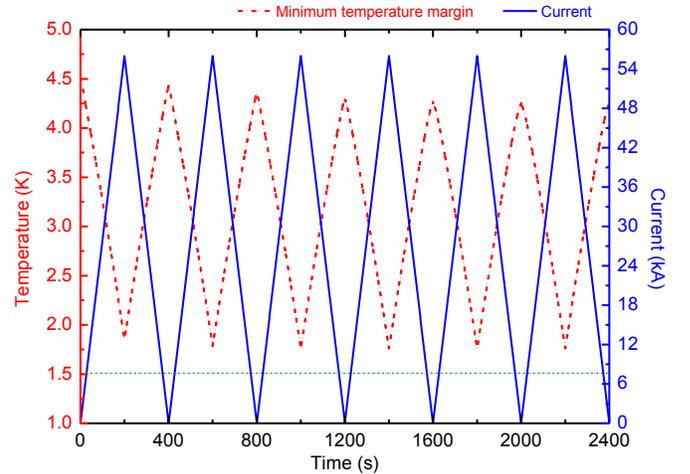

**Figure 11.** Minimum temperature margin and current evolution as functions of time for the B1 channel.

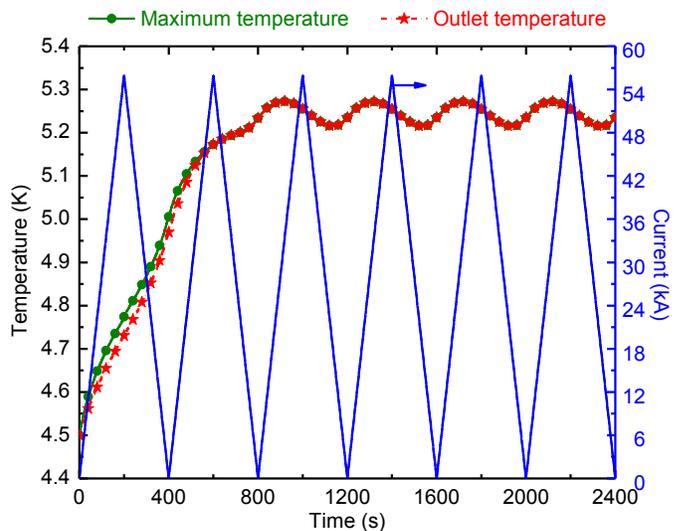

**Figure 12.** Maximum cable temperature, outlet temperature and current evolution as functions of time for the A1 channel.

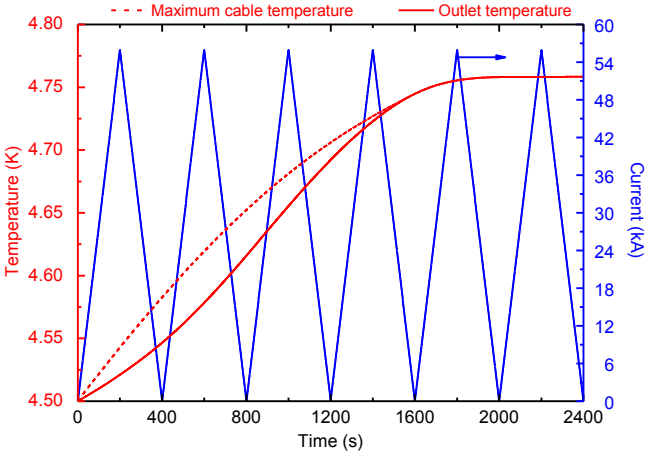

**Figure 13**. Maximum temperature, outlet temperature and current evolution as functions of time for the B1 channel.

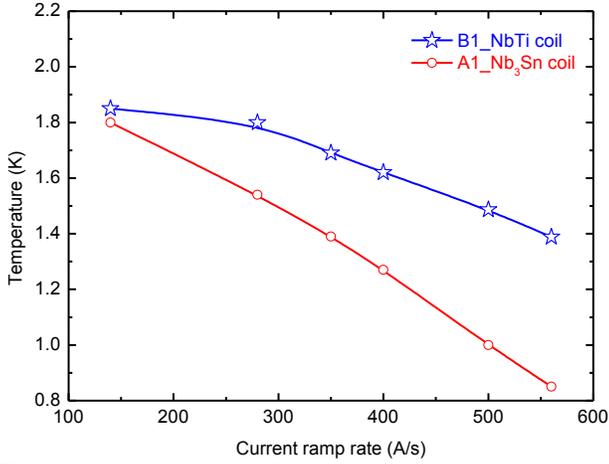

**Figure 14.** The lowest value of the minimum temperature margin vs current ramp rate for cyclic operation.

### D. Quench analysis of the superconducting magnet

The superconducting magnet will store a large magnetic energy of about 436.6 MJ at 56 kA. In case of a quench occurs, the large magnetic energy should be timely discharged into external dump resistors against overheating and voltage breakdown which might be caused by some failure in the superconducting coils [6, 74-80]. Therefore, effective quench detection and quench protection should be taken to protect the superconducting coils from damaging and to prevent the inopportune fast discharges [81]. The dump resistor of ST-08 stainless steel with positive temperature coefficient was used for accelerating the current decay during a quench.

The adiabatic hot spot temperature criterion and the quasi-1 D Gandalf code were used to analyze the thermal-hydraulic behavior of the CICC during a quench. The heat capacity of the cable only was considered for the adiabatic hot spot temperature criterion with maximum hot spot temperature of 250 K. The maximum hot spot temperature is limited to 150 K for the 1-D Gandalf code. Quench can be taken place at any conductor of the inner and outer coils of the superconducting magnet. Therefore, the quench behavior of the two coils should be analyzed respectively. The A1 and B1 channels were selected to model their quench propagation behavior because the temperature margin is the lowest in comparison with the other coil for the Nb₃Sn coil and NbTi coil.

According to the adiabatic hot spot temperature criterion, the holding time can be evaluated. The heat balance equation of the cable for the adiabatic hot spot temperature criterion can be described as follows,

$$J_{Cu}^2 (t + \frac{t_D}{2}) = Z(T_f) \tag{6}$$

$$Z(T_f) = \int_{T_b}^{T_f} (\frac{A_{NonCu} C_{NonCu} \rho_{NonCu}}{A_{Cu} \eta_{Cu}} + \frac{C_{Cu} \rho_{Cu}}{\eta_{Cu}}) dT \tag{7}$$

$$t = t_p + t_h + t_{cb} \tag{8}$$

where $t_p$ is the quench propagation time to threshold voltage from quench initiation, $t_h$ is the quench holding time, and $t_{cb}$ is the time to open the breaker.

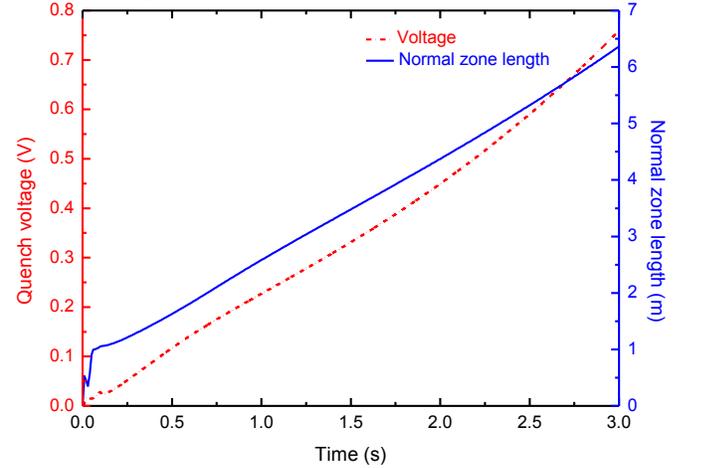

**Figure 15**. Normal zone length and quench voltage evolution as functions of time for the A1 channel.

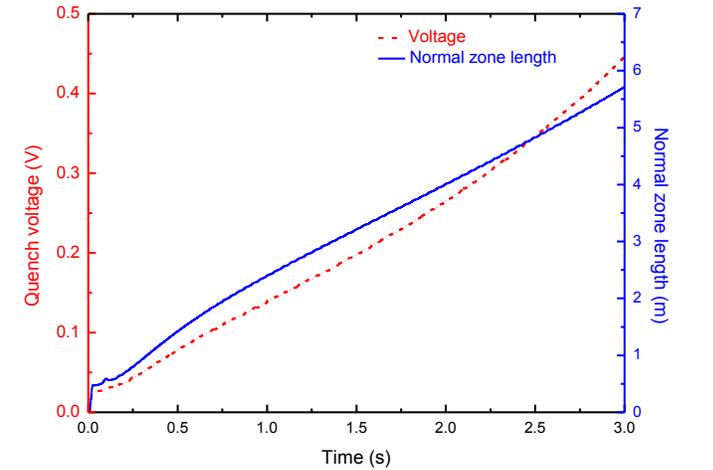

**Figure 16.** Normal zone length and quench voltage evolution as functions of time for the B1 channel.

By using the formula as shown above, we can obtain the maximum holding time. The total equivalent thermal time constant of the A1 channel is about 9.8 s. The quench voltage evolution as a function of time before it was introduced into the external dump resistor was calculated with the 1-D GANDALF code. To initiate a quench, a square heat pulse of (1 m, 0.1 s) was introduced to drive the superconducting coil into the resistive state from the superconducting state. The quench energy adopted is about 2 times the energy needed to initiate a propagating quench. Figure 15 shows the normal zone length and quench voltage as functions of time for the A1 channel. It takes about 1.80 s to reach the quench voltage of 0.4 V. The time for opening the circuit breaker is about 0.5 s. The equivalent thermal discharge time constant of the A1 channel is selected as 2.8 s. The maximum holding time can be obtained as 6.1 s by taking into account the threshold voltage of 0.4 V if the quench originated from the A1 channel.

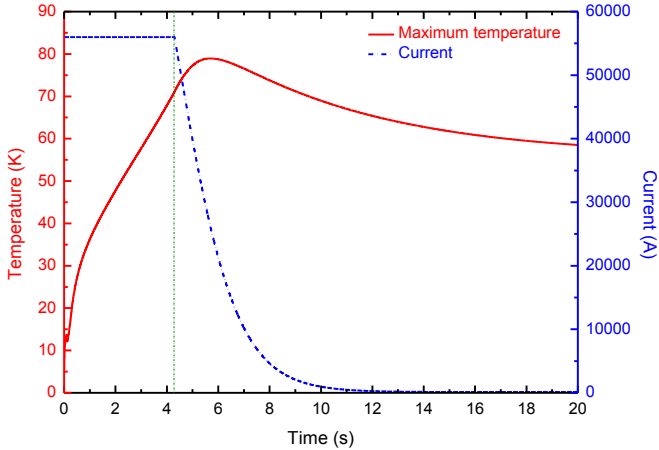

**Figure 17**. Hot spot temperature evolution as a function of time for the A1 channel.

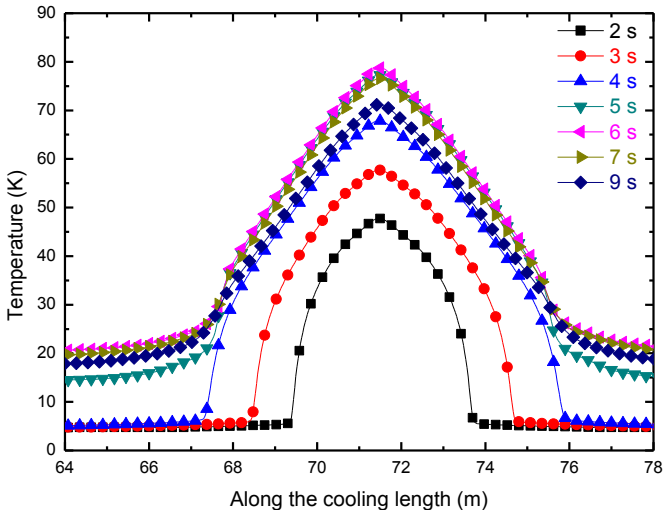

**Figure 18**. Cable temperature evolution along the cooling length as a function of time for the A1 channel.

The total equivalent thermal time constant of the B1 channel is about 8.2 s. Figure 16 shows the normal zone length and quench voltage as functions of time for the B1 channel. It takes about 2.78 s to reach the quench voltage of 0.4 V. The maximum holding time can be obtained as 3.52 s by taking into account the threshold voltage of 0.4 V and the equivalent thermal discharge time constant of 2.8 s if the quench originated from the B1 channel. The threshold voltage and holding time can be designed as 0.4 V and 2.0 s respectively.

Figure 17 shows the hot spot temperature evolution as a function of time for the holding time of 2.0 s, initial disturbance length of 1 m and quench threshold voltage of 0.4 V for the A1 channel. Figure 18 shows the cable temperature evolution along the cooling length as a function of time for the A1 channel. It is shown that the maximum cable temperature is about 78.9 K, which is below the ITER design criterion on the hot spot temperature with the 1D thermal–hydraulic model. Figure 19 shows helium pressure evolution as a function of time for the holding time of 2.0 s, initial disturbance length of 1 m and quench threshold voltage of 0.4 V for the A1 channel. Figure 20 shows the helium pressure evolution along the cooling length as a function of time for different times. The maximum helium pressure is about 4.2 MPa at 5.9 s, which is well below the ITER design criterion of 25 MPa.

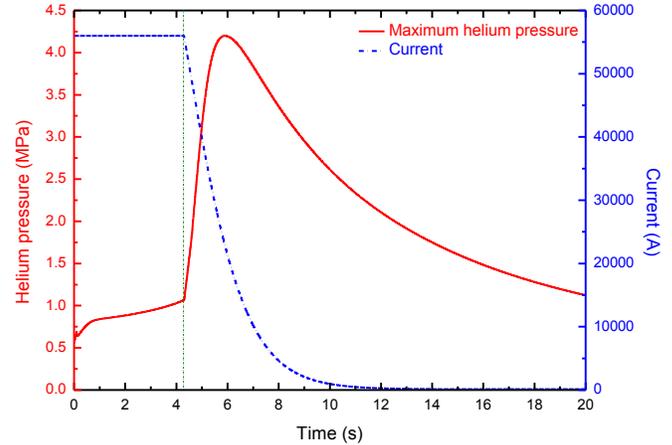

**Figure 19**. Maximum helium pressure evolution as a function of time for the A1 channel.

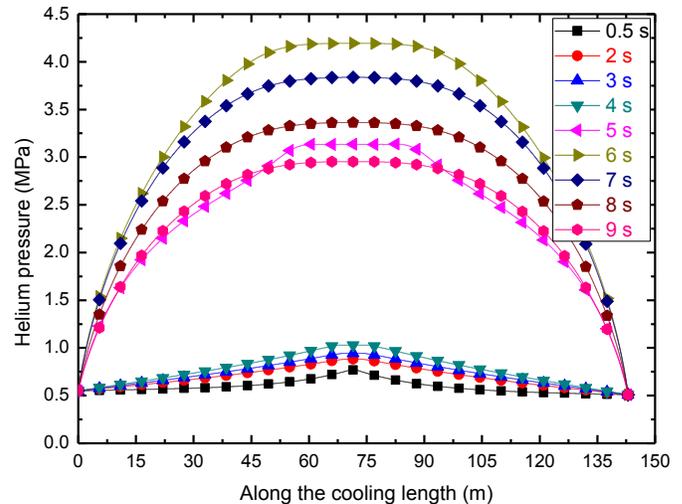

**Figure 20**. Helium pressure evolution along the cooling length as a function of time for the A1 channel.

The parametric analysis has been performed to evaluate the sensitivity of the maximum cable temperature and the maximum helium pressure to the holding time, the initial disturbance length, and the threshold voltage for the Nb$_3$Sn coil. The analysis results are shown in tables 8–10. It is shown that the threshold voltage and the holding time have a negligible impact on the maximum helium pressure while the initial disturbance length has a substantial impact on the maximum helium pressure. It is shown that the maximum helium pressure is about 7.22 MPa for the threshold voltage of 0.4 V, the holding time of 2.0 s and the disturbance length of 10.0 m. The calculated results also show that the maximum cable temperature increase with increasing the holding time and threshold voltage. The maximum cable temperature is about 106.4 K for the threshold voltage of 0.4 V, the holding time of 4.0 s and the disturbance length of 1.0 m. However, the disturbance length has the opposite effects on the maximum cable temperature.

**Table 8.** Sensitivity of maximum cable temperature and maximum helium pressure to disturbance length for the A1 channel.

| Disturbance length (m) | Max. cable temperature (K) | Max. helium pressure (MPa) |
|---|---|---|
| 0.5 | 84.0 | 4.17 |
| 1.0 (Ref.) | 78.9 | 4.18 |
| 10.0 | 58.0 | 7.22 |

**Table 9.** Sensitivity of maximum cable temperature and maximum helium pressure to holing time for the A1 channel.

| Holding time (s) | Max. cable temperature (K) | Max. helium pressure (MPa) |
|---|---|---|
| 2.0 (Ref.) | 78.9 | 4.18 |
| 2.5 | 85.0 | 4.19 |
| 3.0 | 91.6 | 4.22 |
| 3.5 | 98.7 | 4.24 |
| 4.0 | 106.4 | 4.27 |

**Table 10.** Sensitivity of maximum cable temperature and maximum helium pressure to threshold voltage for the A1 channel.

| Threshold voltage (V) | Max. cable temperature (K) | Max. helium pressure (MPa) |
|---|---|---|
| 0.3 | 74.3 | 4.17 |
| 0.4(Ref.) | 78.9 | 4.18 |
| 0.45 | 81.3 | 4.20 |
| 0.50 | 83.8 | 4.23 |

Figures 21 and 22 show the maximum cable temperature and maximum helium pressure evolution as functions of time for the holding time of 2.0 s, initial disturbance length of 1 m and quench threshold voltage of 0.4 V for the B1 channel. Figures 23 and 24 show the cable temperature and helium pressure evolution along the cooling length as functions of time for the B1 channel. It is shown that the hot spot temperature of the NbTi cable is about 76.7 K, which is well below the ITER design criterion on the hot spot temperature with the 1D thermal–hydraulic model. The maximum helium pressure is about 1.66 MPa at 7.1 s, which is well below the ITER design criterion of 25 MPa.

The parametric analysis has also been performed to evaluate the sensitivity of the maximum cable temperature and the maximum helium pressure to the holding time, the initial disturbance length, and the threshold voltage for the B1 channel. The results are shown in tables 11–13. The sensitivity of the maximum cable temperature and maximum helium pressure to the holding time, the initial disturbance length, and the threshold voltage detection for the B1 channel was similar to those achieved for the A1 channel. The maximum cable temperature is about 111.1 K for the threshold voltage of 0.4 V, the holding time of 4.0 s and the disturbance length of 1.0 m. Therefore, the threshold voltage of 0.4 V and holding time of 2.0 s are selected to safely protect the superconducting magnet.

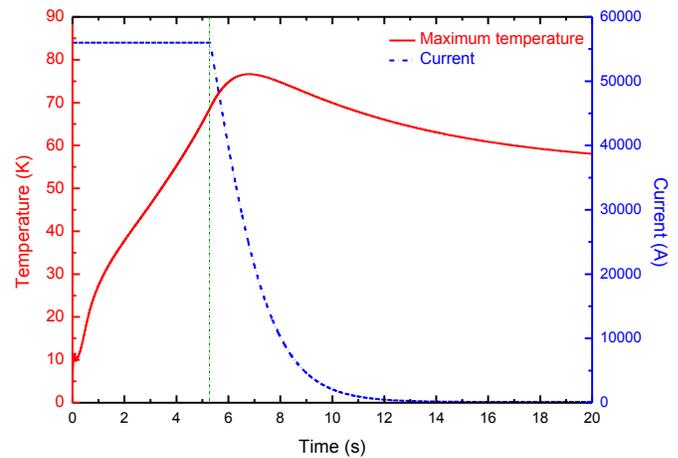

**Figure 21.** Hot spot temperature evolution as a function of time for the B1 channel.

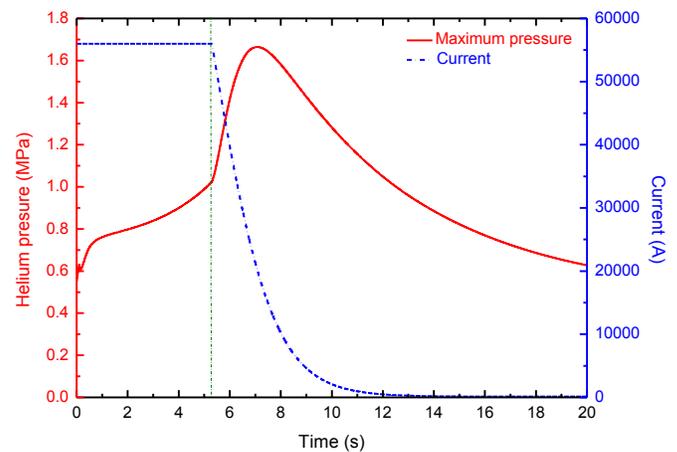

**Figure 22.** Maximum helium pressure evolution as a function of time for the B1 channel.

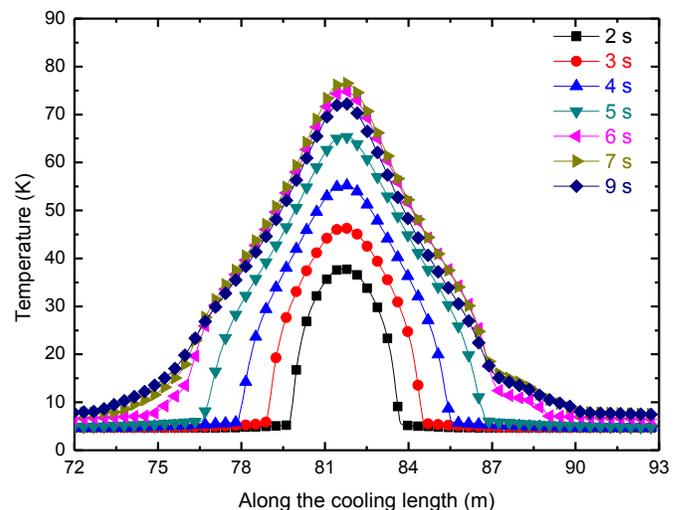

**Figure 23.** Cable temperature evolution along the cooling length as a function of time for the B1 channel.

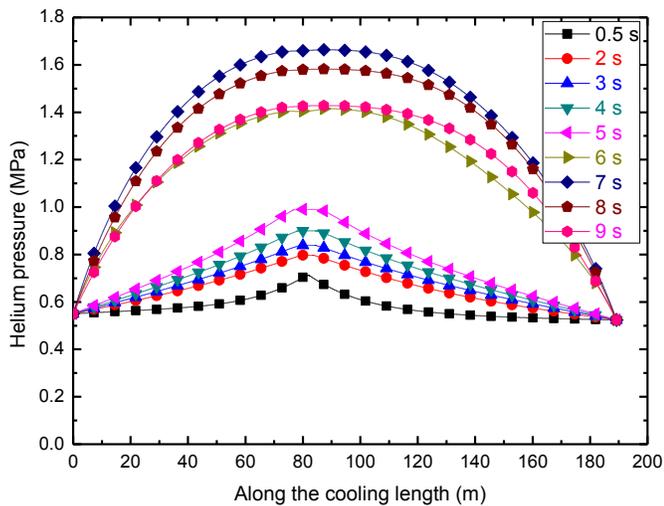

**Figure 24.** Helium pressure evolution along the cooling length as a function of time for the B1 channel.

**Table 11.** Sensitivity of maximum cable temperature and maximum helium pressure to disturbance length for the B1 channel.

| Disturbance length (m) | Max. cable temperature (K) | Max. helium pressure (MPa) |
|---|---|---|
| 0.5 (Ref.) | 82.8 | 1.63 |
| 1.0 | 76.6 | 1.66 |
| 10.0 | 50.3 | 2.93 |

**Table 12.** Sensitivity of maximum cable temperature and maximum helium pressure to holing time for the B1 channel.

| Holding time (s) | Max. cable temperature (K) | Max. helium pressure (MPa) |
|---|---|---|
| 2.0 (Ref.) | 76.6 | 1.66 |
| 2.5 | 83.9 | 1.70 |
| 3.0 | 92.0 | 1.75 |
| 3.5 | 101.0 | 1.81 |
| 4.0 | 111.1 | 1.90 |

**Table 13.** Sensitivity of maximum cable temperature and maximum helium pressure to threshold voltage for the B1 channel.

| Threshold voltage (V) | Max. cable temperature (K) | Max. helium pressure (MPa) |
|---|---|---|
| 0.3 | 68.3 | 1.65 |
| 0.4(Ref.) | 76.6 | 1.66 |
| 0.45 | 82.3 | 1.68 |
| 0.50 | 82.4 | 1.69 |

## VII. CONCLUSION

The design of the superconducting test facility for the CFETR magnet system have been described in this paper. The detailed analysis of the electromagnetic, stress and thermal-hydraulic behavior for the superconducting magnet system is performed. The stress analysis of the superconducting magnet under normal operation conditions is performed by using the 2D finite element method. The stress level of the jacket and insulation for the superconducting magnet satisfies the ITER design criteria. The thermal-hydraulic analysis on temperature margin of the superconducting magnet shows that there are large minimum temperature margins during normal operation. The quench analysis of the superconducting magnet shows that the hot spot temperature and the maximum helium pressure are within the design criteria.

### ACKNOWLEDGMENTS

This work was supported in part by the National Natural Science Foundation of China under Grant No. 51406215, the Anhui Provincial Natural Science Foundation under Grant No. 1408085QE90, and by the National Magnetic Confinement Fusion Program of China (Grant Nos 2014GB106000, 2014GB106003, and 2014GB105002). The authors would like to thank Dr. Wei Tong at High Magnetic Field Laboratory, Chinese Academy of Sciences for his helpful discussion. The views and opinions expressed herein do not necessarily reflect those of the EAST and CFETR Organization.